\documentclass[showpacs,preprintnumbers,amsmath,amssymb]{revtex4}

\pdfoutput=1

\usepackage{amssymb,graphicx,subfigure,url,amsmath,rotating}

\usepackage{dcolumn}
\usepackage{bm}

\usepackage{wasysym}

\begin{document}


\title{Implementation of Glider Guns in the Light-Sensitive Belousov-Zhabotinsky Medium}

\author{Ben de Lacy Costello}
\email{ben.delacycostello@uwe.ac.uk}
\affiliation{Centre for Research in Analytical, Material and Sensor Sciences,  University of the West of England, Bristol, BS16 1QY}

\author{Rita Toth}
\email{rita.toth@uwe.ac.uk}
\affiliation{Centre for Research in Analytical, Material and Sensor Sciences,  University of the West of England, Bristol, BS16 1QY}

\author{Christopher Stone}
\email{christopher3.stone@uwe.ac.uk}
\affiliation{Department of Computer Science, University of the West of England, Bristol, BS16 1QY}

\author{Andrew Adamatzky}
\email{andrew.adamatzky@uwe.ac.uk}
\affiliation{Department of Computer Science, University of the West of England, Bristol, BS16 1QY}

\author{Larry Bull}
\email{larry.bull@uwe.ac.uk}
\affiliation{Department of Computer Science, University of the West of England, Bristol, BS16 1QY}

\begin{abstract}
In cellular automata models a glider gun is an oscillating pattern of non-quiescent states
that periodically emits traveling localizations (gliders). The glider streams can be combined to
construct functionally complete systems of logical gates and thus realize universal computation.
The glider gun is the only means of ensuring the negation operation without additional external input
and therefore is an essential component of a collision-based computing circuit. We demonstrate the existence of
glider gun like structures in both experimental and numerical studies of an excitable chemical system -- the light-sensitive Belousov-Zhabotinsky reaction. These discoveries could provide the basis for future designs of collision-based reaction-diffusion computers.
\end{abstract}

\pacs{05.45.-a; 82.40.Ck; 89.75.Kd, 89.75.Fb, 89.20.Ff}


\maketitle

\section{Introduction}
\label{intro}

The field of unconventional computing involves the design of novel computing devices based on principles of information processing in chemical, physical
and biological systems~\cite{adamatzky_teuscher_2006,adamatzky_2007,akl_2007}. Experimental prototypes include reaction-diffusion processors~\cite{adamatzky_delacycostello_asai_2005, Adamatzky_Phys_Rev, toth_1995, steinbock, Gorecki_2003, Gorecki_phys_rev, Gorecki_2005, Yoshikawa_2001, Yoshikawa_2004, Yoshikawa_2003},
extended analog computers~\cite{mills_2008}, micro-fluidic circuits~\cite{fuerstman_2003},
and plasmodium computers~\cite{nakagaki_2001a,shirakawadelone}.
These computing substrates utilize propagating patterns that encode for information, e.g., chemical waves in reaction-diffusion processors. Every micro-volume can act as both logical gate and wire meaning computation can be performed simultaneously at many points in the reactor. Therefore, it is counterintuitive to impose significant levels of geometrical constraint such as channels to direct information transfer. Instead, the aim is to utilize the natural dynamics of the system to implement computation. The theory of collision-based computing provides a rigorous
computational framework to enable this type of implementation~\cite{adamatzky_cbc}.

Collision-based computing was developed from pioneering results on the
computational universality of the Game of Life \cite{berlekamp_1982}, the billiard-ball model and conservative logic \cite{fredkin_toffoli_1982} and
their implementation in cellular-automaton (CA) models \cite{margolus_1984}.

Collision-based computers employ compact mobile patterns (localizations) to encode elementary units of information.
Truth values of logical variables are given by the presence or absence of localizations or alteration of localization type. The localizations travel in space and implement
logical gates when they collide. There is no imposed architecture such as stationary wires, but instead trajectories of localizations constitute transient wires. The benefits are a hugely adaptable, massively parallel computing medium.

A wide range of physical media are capable of supporting mobile localizations. These include solitons, defects
in tubulin microtubules, excitons in Scheibe aggregates and breathers in
polymer chains~\cite{adamatzky_cbc}. Theoretically, these phenomena should be candidates for implementing collision-based circuits, but in practice they are not easily observed or controlled. A relatively amenable experimental substrate based on the light-sensitive Belousov-Zhabotinsky (BZ) medium~\cite{sendina_2001} was shown to be capable of implementing collision-based gates~\cite{adamatzky_2004,ben_2005,adamatzky_delacycostello_2007,toth_chaos}. The sub-excitable phase of this reaction supports short-lived propagating localizations in the form of unbounded wave fragments that behave like
quasi-particles. The reaction can be made sub-excitable by careful control of the light level. In the light-sensitive BZ reaction increased light decreases excitability by producing Br- ions, the main inhibitor of the reaction. The sub-excitable limit is between the excitable (plane waves propagate) and non-excitable (reaction insensitive to perturbations) phases of the reaction.

In previous experimental and numerical studies~\cite{adamatzky_2004,ben_2005,adamatzky_delacycostello_2007,toth_chaos} the design
of functionally complete logical gates in BZ reactions was demonstrated. An outstanding problem from these studies was how to implement a negation operator without artificially feeding additional wave fragments into the computing medium. In CA models of collision-based computers (see~\cite{adamatzky_cbc}), negation is implemented using autonomous generators of gliders, known as glider guns. This is done by colliding a stream of gliders with other traveling localizations representing data. Until recently there was no experimental proof that glider gun analogs existed in BZ systems. This paper presents new results showing the existence of glider gun analogs in both experimental and numerical studies of the BZ reaction. We also show that the proposed mechanism of glider gun formation in certain CA systems has close parallels with the formation of glider guns in real chemical systems.

The paper is structured as follows. In Sect.~\ref{gliders} we summarize results of glider gun formation in two distinct CA models and propose a mechanism of glider generation by glider
guns. In Sect.~\ref{setup} we describe the experimental setup and numerical parameters used to implement glider guns in the BZ reaction. In Sect.~\ref{mechanism} we discuss the process of glider gun formation in heterogeneous BZ systems. Sect.~\ref{BZguns} describes several examples of glider guns implemented in numerical studies of the BZ reaction, including a computational interpretation of the results. The paper finishes with a section on approaches to obtaining glider guns in architectureless homogeneous BZ systems (Sect.~\ref{homogeneous media}).

\section{Glider gun formation in cellular automata models}
\label{gliders}

We propose that
{\it gliders in the following CA models are produced from interactions of spiral wave-like patterns that constitute the body of the gun and that these are analogous to  wave fragments in a chemical reaction-diffusion medium}. A comprehensive proof of this proposition is still somewhat distant as it would require exhaustive
analysis of glider guns formed by all possible cell-state transition rules. The CA lattices $2^+$-medium~\cite{adamatzky_2001} and Spiral Rule~\cite{adamatzky_wuensche_delacycostello,adamatzky_wuensche_2007} automata demonstrate our reasons for suggesting this mechanism.

\subsection{Cellular automaton model of sub-excitable chemical media}

The $2^+$-medium cellular automaton is a two-dimensional orthogonal array of cells where each cell has three possible states: resting, excited and
refractory. A cell's state is updated in discrete time depending on the states of its closest neighbors.
A resting cell becomes excited if two neighboring cells are in the excited state. Excited cells become refractory and refractory
cells adopt a resting state at the next time step . The $2^+$-medium is an excitable discrete system capable of universal computation~\cite{adamatzky_2001}. In previous work, we implemented few-bit adders and multipliers in the $2^+$-medium~\cite{liang_adder,liang_multiplier}.

\begin{figure}
\centering
\includegraphics[width=0.7\linewidth]{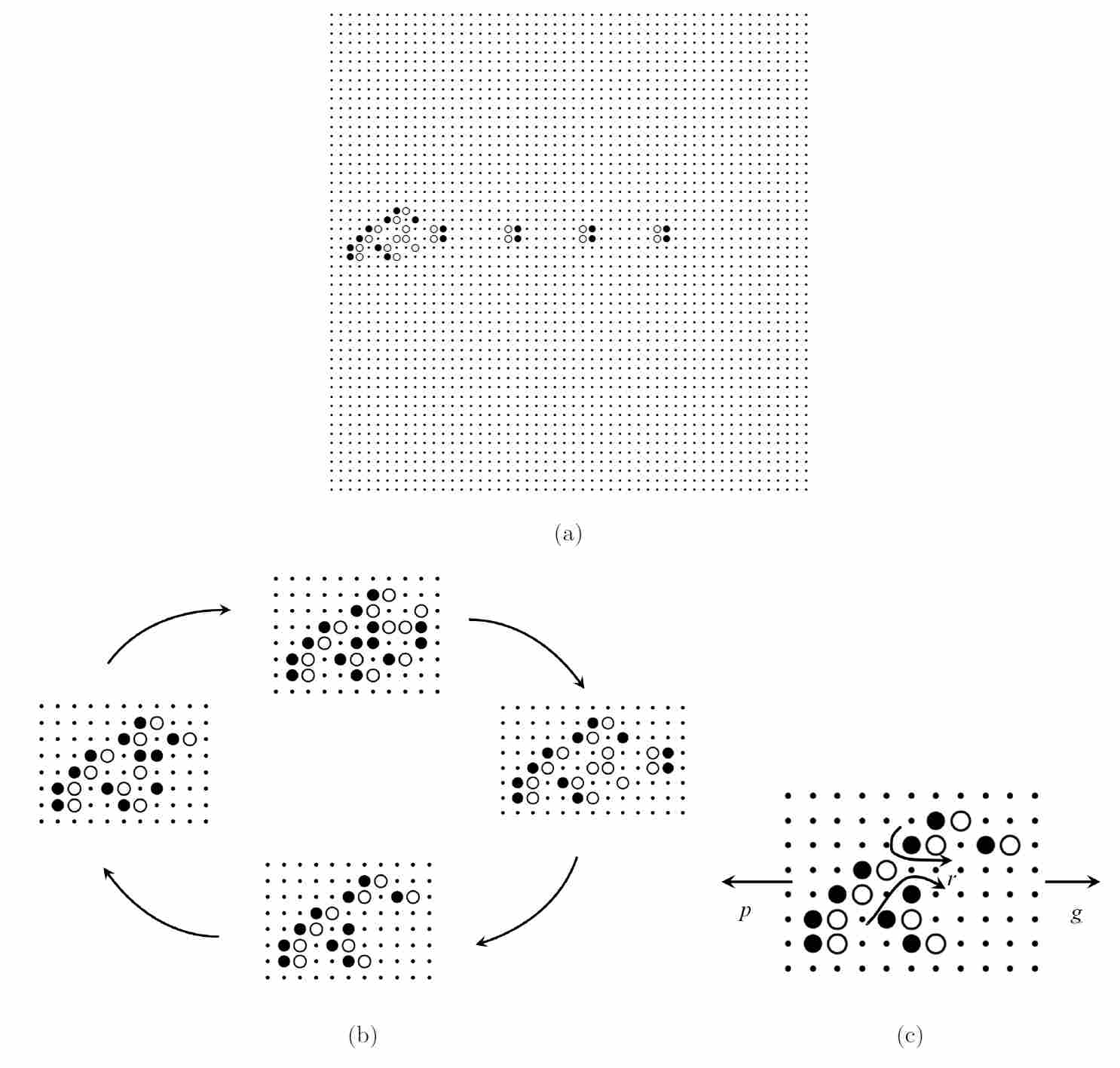}
\caption{Glider gun in the $2^+$-medium.
(a)~The gun travels West and emits gliders (self-localized excitations) traveling East.
(b)~Oscillation cycle of the gun.
(c)~Mechanics of glider generation, $p$ is the gun's velocity vector, $g$ is the direction of the glider stream,
$r$ shows the inward rotation of the wave tips. A solid circle indicates the excited state, an open circle indicates the refractory state and a dot indicates the resting state.}
\label{2medium}
\end{figure}

The simplest mobile localizations (gliders) in the $2^+$-medium are comprised of four non-resting states, a head with two excited states and a tail with two refractory states ($2^+$-particles). Several mobile glider guns (where the body of the gun moves) that emit $2^+$-particles have been identified. To date, no stationary glider guns have been discovered in the $2^+$-medium. A mobile glider gun moving Westwards emitting $2^+$-particles in the opposite direction is shown in Fig.~\ref{2medium}(a).

Whilst traveling, the gun undergoes structural transformations (Fig.~\ref{2medium}(b)). On closer inspection the transformations involve the interaction of two spiral excitation waves within the body of the gun. These spirals collide to form the $2^+$-particle (Fig.~\ref{2medium}(b-c), direction of rotation shown by arrows).

\subsection{Reaction-diffusion based CA model}

\begin{figure}[!htb]
\centering
\includegraphics[width=0.3\linewidth]{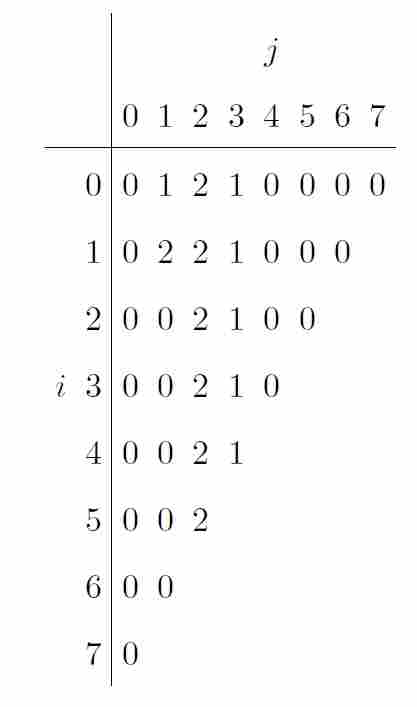}
\caption{
Transition matrix for the Spiral Rule CA~\cite{adamatzky_wuensche_2007}.
}
\label{spiral-rule-transition}
\end{figure}

The Spiral Rule CA~\cite{adamatzky_wuensche_delacycostello,adamatzky_wuensche_2007}, derived from
Wuensche's beehive rule~\cite{wue05}, consists of a hexagonal array of cells, where each cell takes one of three states $\{2, 1, 0\}$. Cells update their state depending on the state of their neighbors. The cell-state transition rules~\cite{adamatzky_wuensche_delacycostello} can be
represented as a matrix where $i$ is the number of neighbors in state 2 and $j$ is the number of neighbors in state 1. Therefore, according to its neighborhood a cell will update to cell-state $M[i][j]$ at the next time step (Fig.~\ref{spiral-rule-transition}).

In~\cite{adamatzky_wuensche_2007}, the Spiral Rule CA was interpreted in terms of an abstract
reaction-diffusion medium where `1' is the activator $A$, `2' is the inhibitor $I$, and `0' is the substrate $S$. Therefore, the transition
matrix (Fig.~\ref{spiral-rule-transition}) represents an abstract chemical reaction. Full details of the quasi-chemical reactions are described in~\cite{adamatzky_wuensche_2007}.

\begin{figure}
\centering
\includegraphics[width=0.7\linewidth]{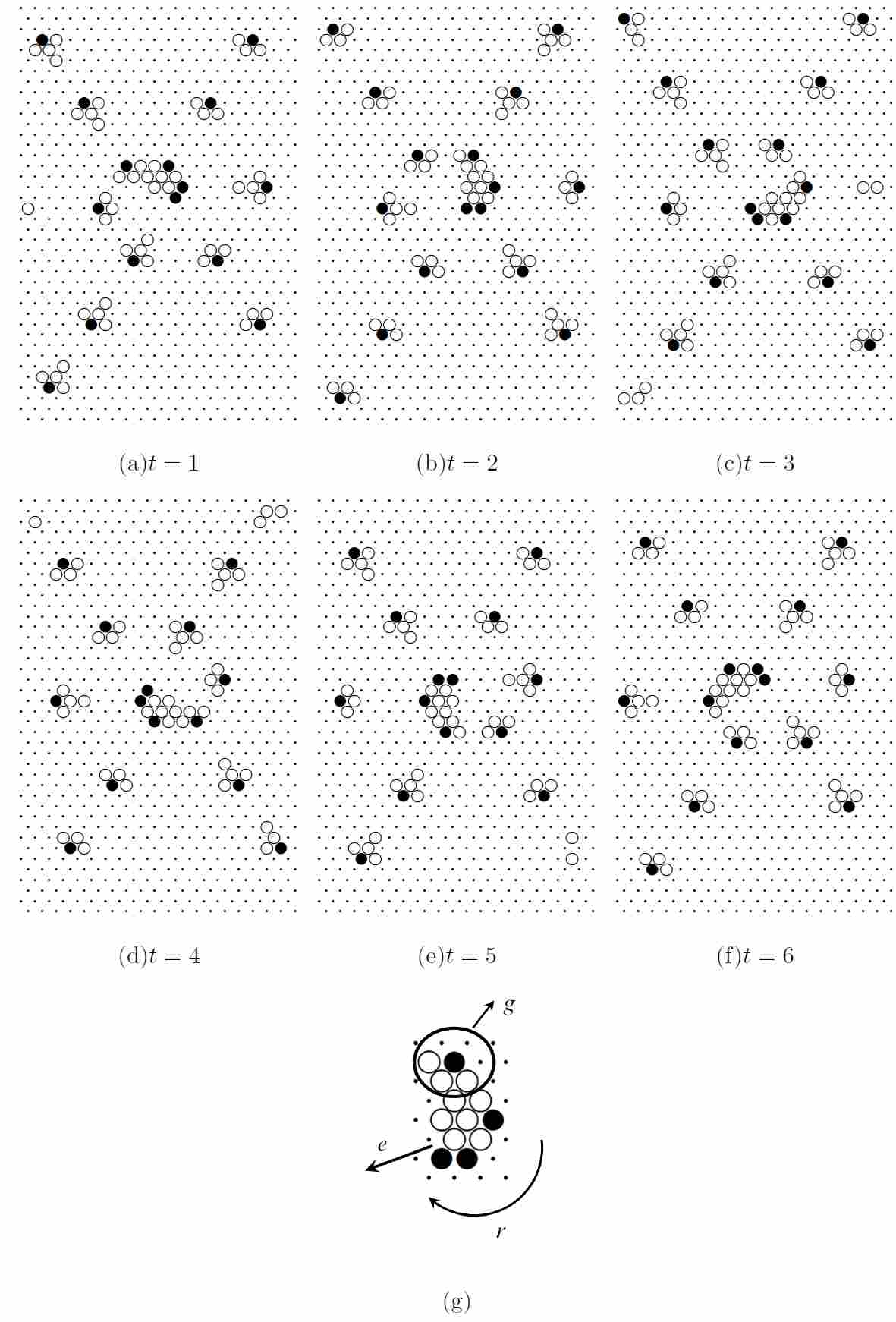}
\caption{Glider gun in the Spiral Rule hexagonal CA: (a)-(f) oscillation cycle of the gun,
(g) mechanics of glider production, $g$ is the velocity vector of a glider traveling North-East, $r$ indicates the
rotation of the spiral wave, $e$ is an excitation force. Activator state is shown by a solid circle, inhibitor state
by an open circle and substrate by a dot.}
\label{spiralhex}
\end{figure}

As shown in Fig.~\ref{spiralhex}(a), a remarkable feature of the Spiral Rule CA is that it exhibits rotating glider guns. The gun is a spiral wave fragment with a few activator states, mainly concentrated at the head of the wave.  We hypothesize that the tail of the spiral
wave, containing one activator state, periodically loses stability and becomes detached. This process forms a propagating localization (glider) with one activator state and three inhibitor
states (Fig.~\ref{spiralhex}(b)). The tail is regenerated and the mechanism repeated, resulting in streams of gliders.

\section{Glider guns in the light-sensitive BZ reaction}
\label{setup}

\subsection{Experimental}

\begin{figure}
\centering
\includegraphics[width=0.7\linewidth]{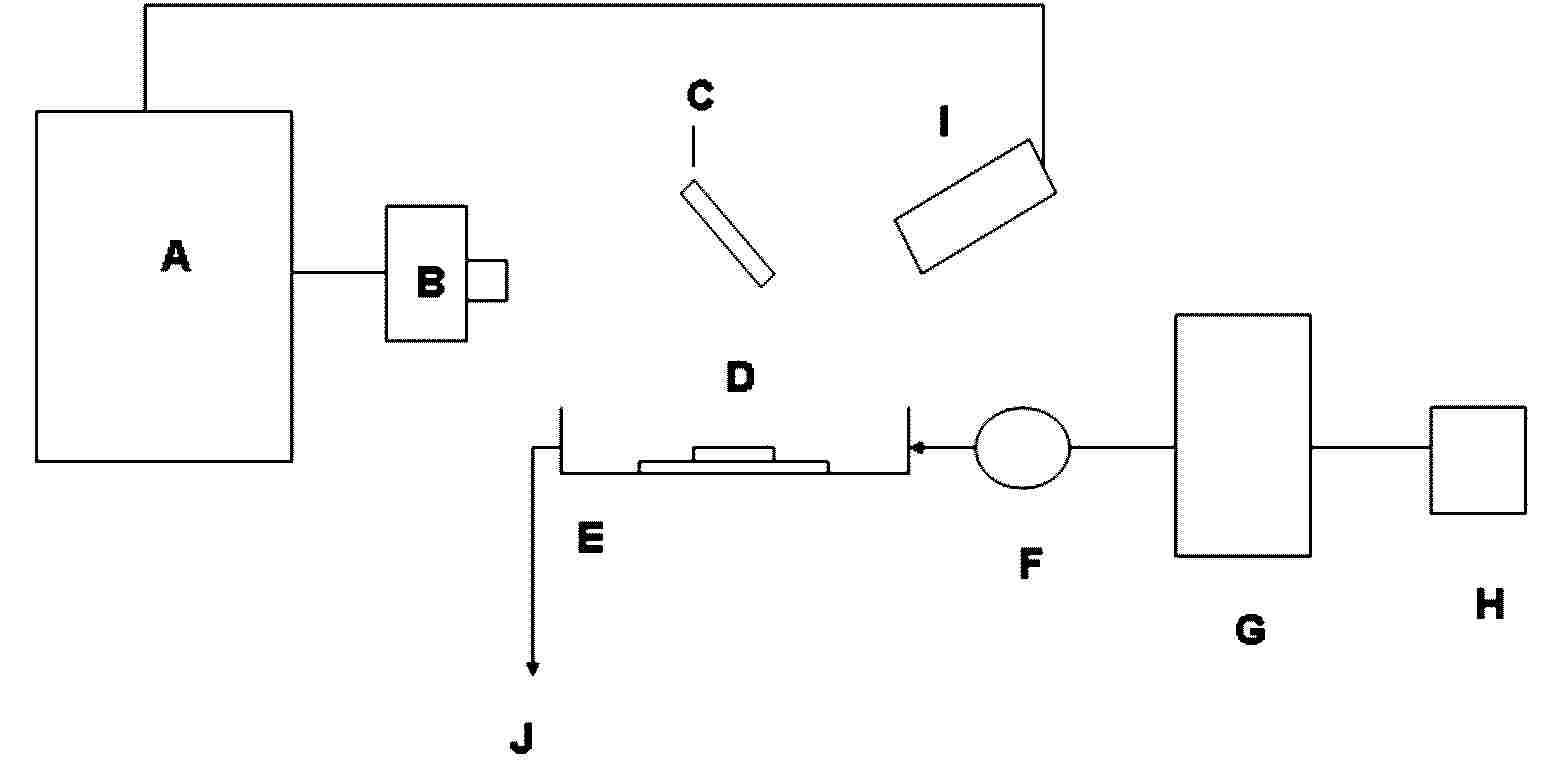}
\caption{Light-sensitive catalyst (Ru(bpy)$_3^{2+}$) [0.004M] loaded silica gel (D) (\cite{wang_1999}) immersed in thermostated (G) Petri dish (E) containing catalyst-free BZ reagents [NaBrO$_3$]=0.36~M, [CH$_2$(COOH)$_2$]=0.0825~M, [H$_2$SO$_4$]=0.18~M, [BrMA]=0.165~M. At this solution composition the reaction is oscillatory in the dark. A peristaltic pump (F) continuously feeds the reactor with solution and removes effluent (J). The reaction solution (H) is in an ice bath. The checkerboard is generated by computer (A) and projected (B) through a 455~nm narrow bandpass interference filter, lens pair and mirror assembly (C). Images are captured by a digital camera (I) and processed to identify chemical wave activity.}
\label{experimentalsetup}
\end{figure}

A light-sensitive BZ reaction~\cite{gaspar_1983} was used, as detailed in Fig.~\ref{experimentalsetup}. A controllable network of excitable (low light 0.394~mW$\cdot$cm$^{-2}$) and non-excitable (high light 9.97~mW$\cdot$cm$^{-2}$) cells arranged in a checkerboard pattern was projected onto the gel surface~\cite{schebesch_1998,sendina_1998,toth_2008}.

\subsection{Numerical}

The Oregonator model modified to account for the photochemistry was used~\cite{Kadar}.

\begin{displaymath}
\frac{\partial u}{\partial t}=\frac{1}{\epsilon}\left[u-u^2-(fv+\Phi)\frac{u-q}{u+q}\right]+D_u\nabla^2u, \quad
\frac{\partial v}{\partial t}=u-v
\end{displaymath}

Model parameters $u$ and $v$ are the dimensionless concentrations of the activator and the oxidized form of the catalyst, respectively. $q=0.0002$ and $\epsilon$ are scaling parameters, $f$ is a stoichiometric coefficient.  $D_u=1$ is the diffusion coefficient of $u$, while $D_v=0$ since $v$ is immobilized
in the gel. The system was integrated using the Euler method with a five-node Laplacian operator ($\nabla^2$), time step $\Delta t=0.001$ and grid point spacing
$\Delta x=0.62$. $\mathit\Phi$ is the rate of production of the photo-induced inhibitor. Low excitability corresponds to high values of $\mathit\Phi$ (equivalent to high light) and vice versa. At $\mathit{\Phi}=0$ the medium is oscillatory. The medium is excitable up to $\mathit{\Phi}=0.034$, weakly-excitable between $\mathit{\Phi}=0.034$ and $\mathit{\Phi}=0.0355$ (the sub-excitable threshold) and non-excitable above $\mathit{\Phi}=0.0355$. The model network consisted of alternating excitable ($\mathit{\Phi_1}$) and non-excitable ($\mathit{\Phi_2}$) squares.
Waves were initiated from an oscillating square ($\mathit{\Phi}=0$) and directed into excitable squares of the network.

\section{Mechanism of glider gun formation in BZ systems}
\label{mechanism}
\begin{figure}
\centering
\includegraphics[width=0.7\linewidth]{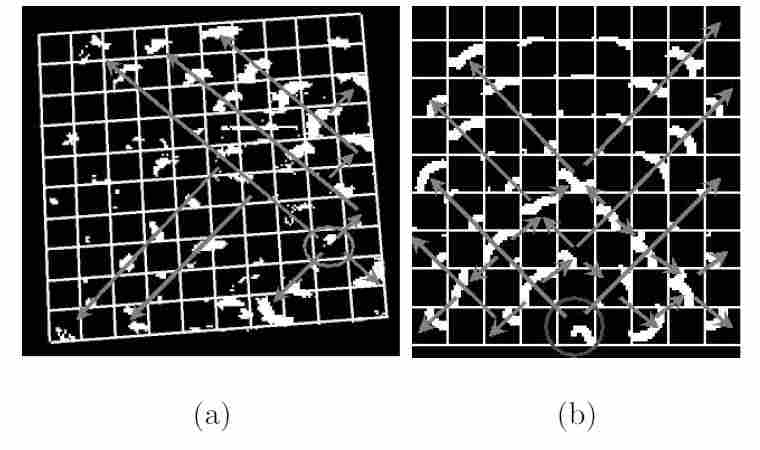}
\caption{Glider guns formed in (a) experimental and (b) numerical studies of the BZ reaction.
Grey arrows show the propagation direction of glider streams. All the streams of gliders come from a single glider gun (spiral rotating within one excitable cell) marked by a circle.
Conditions (a) $\mathit{\Phi_1}=0.394$~mW$\cdot$cm$^{-2}$ and $\mathit{\Phi_2}=9.97$~mW$\cdot$cm$^{-2}$, (b) $\mathit{\Phi_1}=0.003$, $\mathit{\Phi_2}=0.0255$, $\epsilon=0.25$, $q=0.0002$
$f=1.11$, $\Delta x=1.3$. Grid size (a) approximately 22$\times$22mm, (b) 135$\times$135 simulation points.}
\label{exp}
\end{figure}

Glider guns in BZ systems are rotating spiral waves which become trapped within single excitable cells in a network of light (non-excitable) and dark (excitable) cells. Excitable cells have up to four neighboring excitable cells joined by corner to corner junctions. Therefore, streams of gliders (excitation waves) can be emitted in four directions. Examples of glider guns in numerical and experimental BZ systems that emit glider streams in all possible directions are shown in Fig.~\ref{exp}.

Glider guns are formed when spiral waves with trajectories pinned around non-excitable cells degenerate, giving spirals trapped in excitable cells. If waves are input to networks with high excitability fragmented stable waves are formed that travel through all corner to corner junctions of the excitable cells. However, if network excitability is reduced ($\mathit{\Phi_1}$ and $\mathit{\Phi_2}$  moved closer to the sub-excitable limit) spirals form as the fragmented waves become disconnected due to junction failure. These spirals are often complex interacting spirals with their trajectories pinned around non-excitable cells or regions. Therefore, fragments are incident repeatedly at certain junctions and eventually only critical amounts of activator (on the boundary between expansion and annihilation) can cross these junctions. These critical excitation waves form as the refractory period of the junction is only fractionally longer than the period of rotation of the original spiral(s). Critical excitation waves do not expand rapidly but instead drift across the cell. When they eventually expand they possess high curvature enabling them to curl back to form spirals trapped within excitable cells. Wave curvature increases with excitability~\cite{Mihaliuk} and the core of a spiral wave decreases~\cite{Zykov}. Therefore, excitable cells in the network must possess a critical size and excitability in order for these spirals to form. If excitability is too high the initial waves are stable and no spirals form in the network. If excitability is too low, spirals cannot form in excitable cells because wave curvature is too low.

Spirals trapped within excitable cells have a shorter period of rotation than the parent spirals. Over time a number of these spirals form
at various points in the network. This is because these spirals emit a proliferation of wave fragments meaning that the original mechanism of formation has an increased chance of being repeated.

The glider guns formed in BZ systems exhibit strong parallels with glider guns formed in the Spiral Rule CA. In this CA a rotating spiral wave emits glider streams in all possible directions (Fig.~\ref{spiralhex}(a)). In the BZ system the interaction between the rotating spiral wave, the cell boundaries and junctions controls the firing of the gun. In Fig.~\ref{exp} the junctions are relatively excitable and therefore, when a glider gun forms, gliders are emitted in all possible directions. So despite the periodic motion of the spiral, a super-critical amount of activator always reaches the junction and stable gliders form.

\begin{figure}
\centering
\includegraphics[height=90mm]{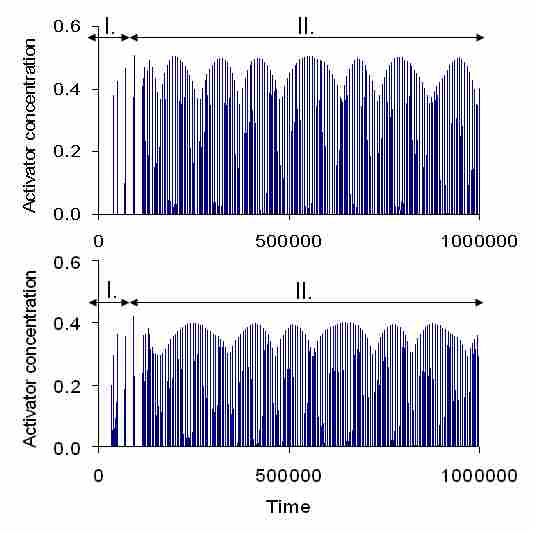}
\caption{Activator concentration measured in numerical studies (at two opposite junctions in a cell) showing the formation of a glider gun. In region (I) fragments incident at the junctions are widely spaced as spiral waves in the network have trajectories pinned around non-excitable cells. In region (II) a spiral rotating in an excitable cell (glider gun) forms and closely spaced fragments are incident at the junctions. The amount of activator at each corner varies periodically because the spiral tip meanders within the cell. This observed periodicity controls the number and direction of glider streams emitted from the cell. Figures show dimensionless time and concentration units.}
\label{activator}
\end{figure}

In experiment, if excitability levels are reduced further then wave transfer is generally unstable because of added heterogeneity (gel imperfections, light dispersion etc.) when compared to numerical studies. In numerical studies we were able to map all excitability levels where glider guns formed. Glider guns that only emit wave fragments in selected directions were observed at lower excitability levels. In some cases glider guns fire intermittently into certain neighboring cells. This
intermittent firing pattern can be explained by the periodic shift in rotation of the spiral (Fig.~\ref{activator}). These `unstable' glider guns are easier to fit
into collision-based computational schemes as stable glider guns completely dominate network dynamics (see Sect.~\ref{BZguns}). However, the fact that both types exist in the model indicates that unstable types may be observed in experiments where heterogeneity can be reduced.

\section{Computing with glider guns in BZ media}
\label{BZguns}

\begin{figure}
\centering
\includegraphics[width=0.7\linewidth]{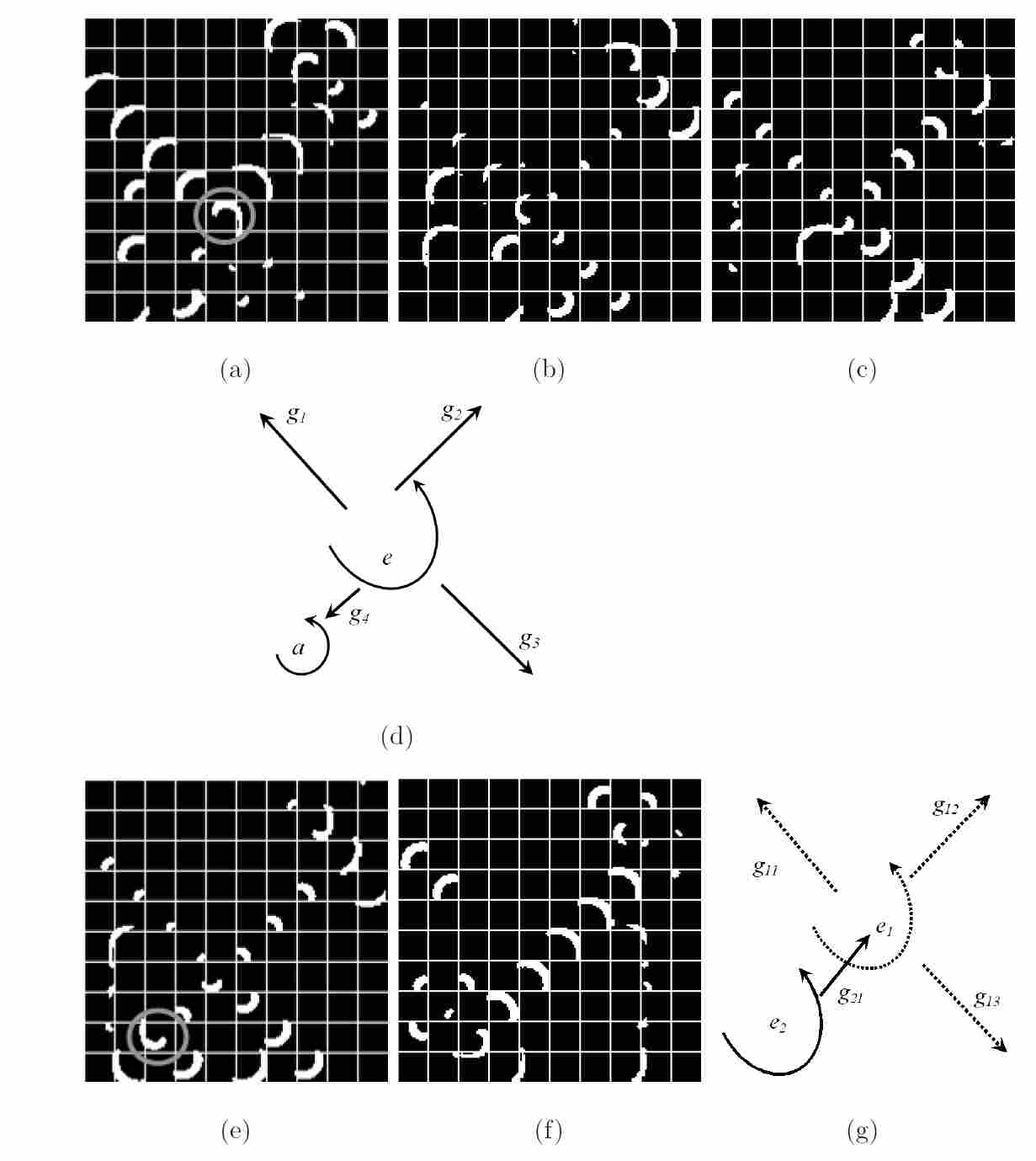}
\caption{(a)--(c)~Excitation dynamics from numerical simulation of the BZ reaction showing cancelation of glider stream. (d)~Reaction scheme showing that glider gun $e$ generates four glider streams $g_1 \cdots g_4$. The unstable stream $g_4$ is canceled via collision with auxiliary spiral
wave $a$. (e) Following on from (c) the excitation dynamics show the formation of a second glider gun $e_2$ (marked by a circle). (f)~Cancelation of the parent glider gun $e_1$ by the daughter glider gun $e_2$ (Reaction scheme shown in (g)). Network excitability $\mathit{\Phi_1}=0.0245$ and $\mathit{\Phi_2}=0.0465$. Model parameters $\epsilon=0.11$, $\Delta x=0.62$. Grid size 180$\times$180 simulation points.}
\label{gun2}
\end{figure}

\begin{figure}
\centering
\includegraphics[width=0.7\linewidth]{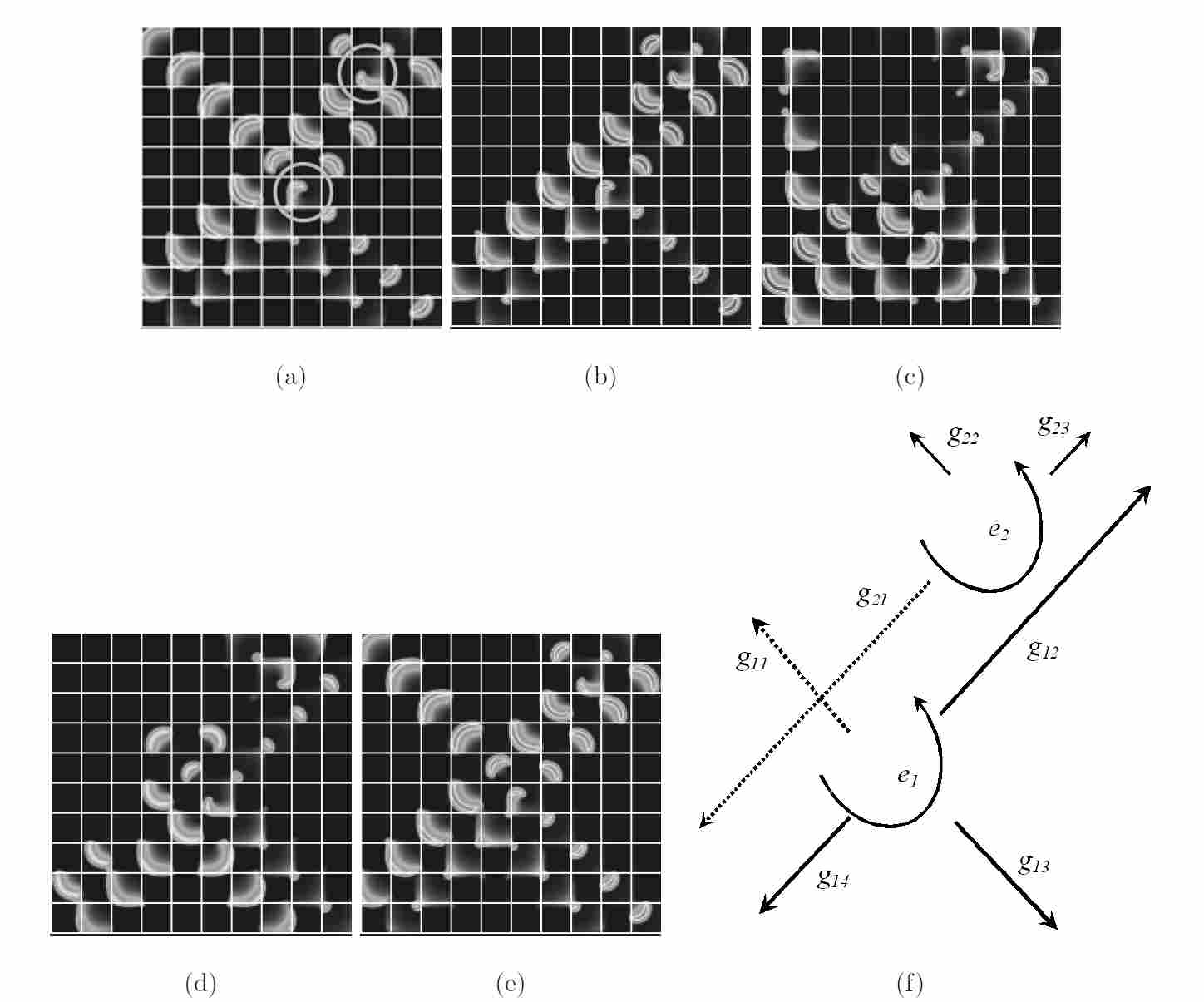}
\caption{(a)--(e)~Excitation dynamics in numerical simulations of the BZ reaction showing the switching cycle between two interacting glider streams. (f)~Schematic representation showing that glider gun $e_1$ initially generates four streams of gliders $g_{11} \cdots g_{14}$ (see (a)). Glider stream $g_{11}$ is temporarily switched off via interaction with stream $g_{21}$ from glider gun $e_2$ (see(b)). Glider stream $g_{11}$ is switched back on again after stream $g_{21}$ is interrupted by interaction with fragments from stream $g_{12}$ (see (c),(d)). Network excitability $\mathit{\Phi_1}=0.0245$ and $\mathit{\Phi_2}=0.0465$. Model parameters $\epsilon=0.11$, $\Delta x=0.62$. Grid size 180$\times$180 simulation points.}
\label{switch}
\end{figure}

\begin{figure}
\centering
\includegraphics[width=0.7\linewidth]{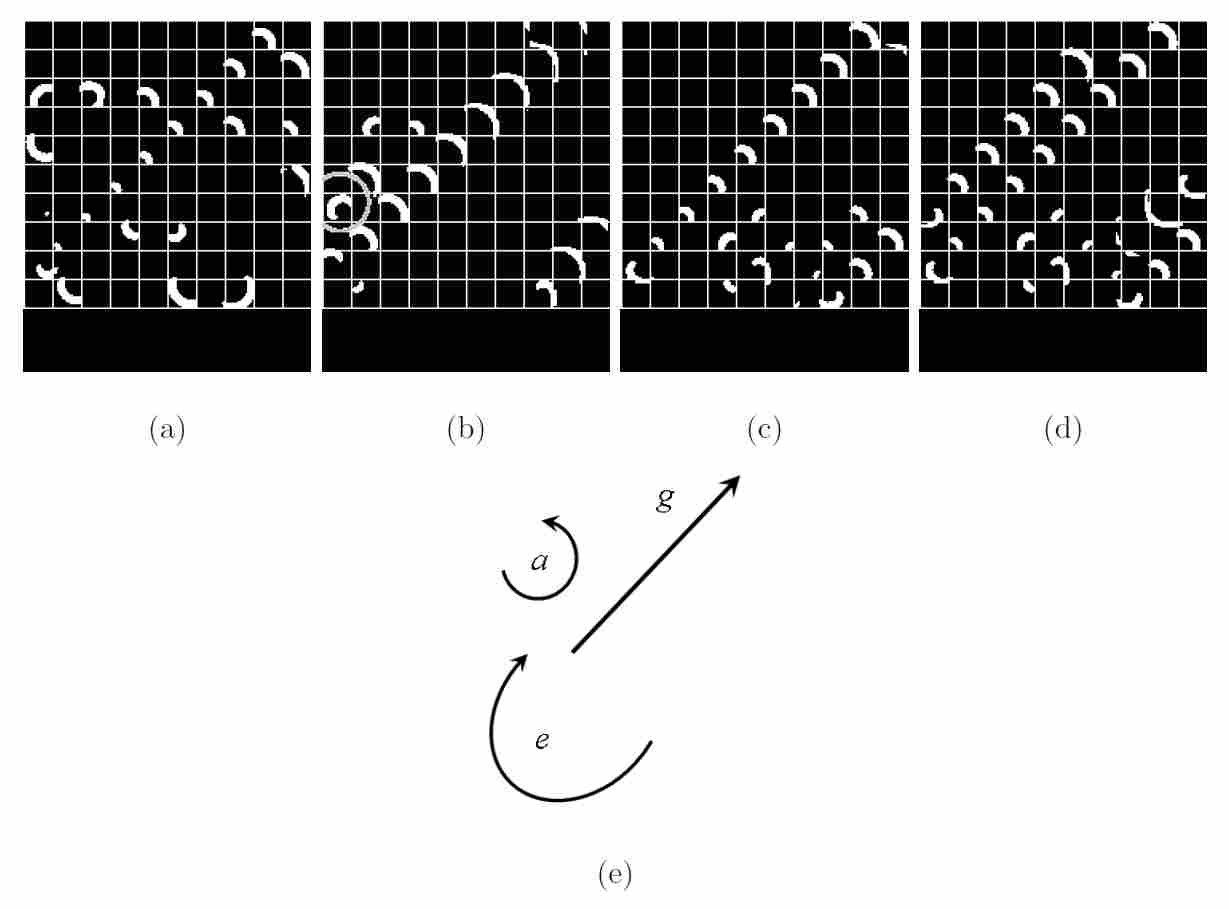}
\caption{Implementing simple memory analogs in numerical studies of the BZ reaction. (a)-(d) Excitation dynamics showing that stable spiral wave $a$ (a) is annihilated by fragments emanating from the intermittent firing of glider gun $e$ (b). This can be interpreted as the stored information being erased thus emptying the memory (c). Another glider gun is formed (d) representing a constant input to the memory (reducing the size of the addressable zone). (e) shows the original reaction scheme where $a$ represents the initial information stored in the memory. The memory unit is formed by the segregation of the reactor by the glider gun $e$ and stream of gliders $g$. Network excitability $\mathit{\Phi_1}=0.0245$ and
 $\mathit{\Phi_2}=0.0465$. Model parameters $\epsilon=0.11$, $\Delta x=0.62$.  Grid size 180$\times$180 simulation points.}
\label{gun3}
\end{figure}

In our numerical studies we have observed glider guns with 2, 3 and 4 stable streams of gliders (Fig.~\ref{gun2},~\ref{switch}). In addition, many glider guns exhibit an unstable intermittent glider stream from
one or more of the `empty' junctions (Fig.~\ref{gun2},~\ref{gun3}). The factors controlling the number of stable and intermittent glider streams are differences in the excitability of the cells and the complex periodic motion of the spirals trapped within excitable cells (Fig.~\ref{activator}).

As mentioned, in CA models gliders play an important role in implementing negation gates and thus ensuring computational universality. Glider guns formed in the BZ reaction can be used for the same purpose. The following reaction schemes show how `data' wave fragments interact with glider streams in numerical studies of the BZ reaction. The work also shows how glider guns and glider streams can be manipulated and annihilated and the results interpreted in terms of computation.

\begin{figure}
\centering
\includegraphics[width=0.7\linewidth]{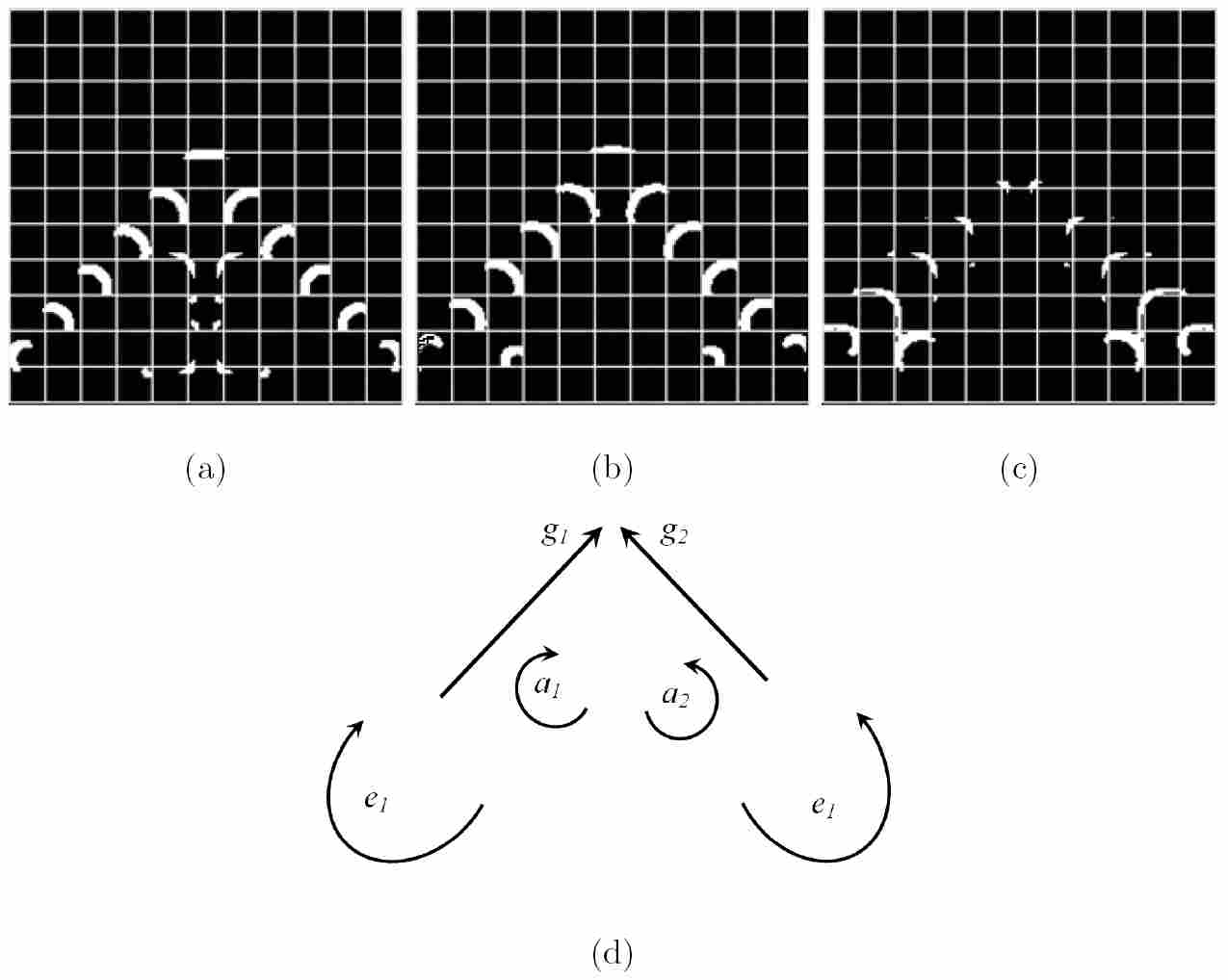}
\caption{(a)--(c) Excitation dynamics in numerical simulations of the BZ medium. Network excitability $\mathit{\Phi_1}=0.0245$ and $\mathit{\Phi_2}=0.0465$. $\epsilon=0.11$, $\Delta x=0.62$ (d) Schematic of glider gun interactions: $e_1$ and $e_2$ are spiral waves (glider guns)
generating glider streams $g_1$ and $g_2$ that interact with auxiliary spiral waves $a_1$ and $a_2$. Grid size 198$\times$198 simulation points.}
\label{gun1}
\end{figure}

Fig.~\ref{gun1} shows the space-time evolution dynamics of two glider guns marked $e_1$ and $e_2$
and auxiliary spiral wave fragments marked $a_1$ and $a_2$ in Fig.~\ref{gun1}(d). The spirals $e_1$ and $e_2$ generate localized  wave fragments (analogs of gliders) traveling North-East (stream $g_1$) and North-West (stream $g_2$). At this network excitability gliders can only cross junctions that are opposite their direction of travel. Compare with Fig.~\ref{exp} where excitability is higher and fragments emanating from the original glider gun cross all junctions in the network. In Fig.~\ref{gun1} the auxiliary spiral waves $a_1$, $a_2$ collide with the glider gun streams and cannot expand further into the network. The glider guns generate a constant output regardless of the presence of auxiliary `data' wave fragments. Interestingly, where glider streams intersect they annihilate. The glider streams segregate the network into zones which is useful from a computational standpoint. This is because multiple reactions with `data' fragments can be realized in parallel.

In CA models of glider guns~\cite{adamatzky_wuensche_2007}, the number of glider streams can be controlled by placing stationary localizations (termed `eaters') at various points in the network. These `eaters' destroy gliders by colliding with them.
In the BZ medium we can employ localized spiral waves to annihilate redundant streams, as in Fig.~\ref{gun2}. Here, the glider gun $e$ in Fig.~\ref{gun2}(d) rotates counterclockwise generating
three stable glider streams. The fourth stream ($g_4$) is annihilated by a spiral wave ($a$) rotating around a non-excitable cell.

When constructing dynamically reconfigurable circuits it is important to be able to remove obsolete glider guns and generate or reposition existing glider guns by using other mobile localizations. This is useful where the bit-range of a circuit should be increased or when switching an arithmetic circuit between an adder and a multiplier. This type of operation has been successfully implemented in CA models~\cite{wue05,adamatzky_wuensche_delacycostello,adamatzky_wuensche_2007}. In CA models,
glider guns are produced in collisions between several gliders. Glider guns can be extinguished via the collision with appropriate timing of a glider at a certain angle. In numerical studies of the BZ reaction we observed the same behavior. Initially, auxiliary spiral fragments and glider streams from $e_1$ combine to form a glider gun $e_2$ (marked by a circle in Fig.~\ref{gun2}(e)). The `daughter' glider gun $e_2$ is positioned in a such a way that stream $g_{21}$ eventually annihilates the parent glider gun $e_1$ (Fig.~\ref{gun2}(f-g)).

Glider streams can be switched on and off periodically (blocked and unblocked) if two multi-streamed glider guns adopt a specific spatial arrangement (Fig.~\ref{switch}). Each of the glider guns ($e_1$,$e_2$)  would emit four streams of wave fragments but stream $g_{12}$ stops all gliders traveling in a South-East direction from $e_2$ (Fig.~\ref{switch}(a)(f)). Glider stream $g_{11}$ is periodically interrupted via interaction with glider stream $g_{21}$. Glider gun $e_2$ is continually firing but the glider stream $g_{21}$ is temporarily interrupted at a point just below the gun due to interaction with glider stream $g_{12}$ (Fig.~\ref{switch}(c)). This gap in stream $g_{21}$ allows stream $g_{11}$ to be switched back {\sc on} (Fig.~\ref{switch}(d),(e)). However, glider stream $g_{11}$ is switched {\sc off} again for $360$ dimensionless time units (t.u.) due to interaction with glider stream $g_{21}$. This switching sequence was observed over many complete cycles. This type of switching circuit observed in the chemical system provides a good example of how complex computational architectures may be achieved.

The switching sequence of wave fragments in streams $g_{11}$ and $g_{21}$ at the point of collision is described as follows:

$$
\begin{array}{l}
g_{11} =  \underbrace{1 \ldots 1}_{\text{ 180 t.u.}} \underbrace{0 \ldots 0}_{\text{ 360 t.u.}} \underbrace{1 \ldots 1}_{\text{ 180 t.u.}} \underbrace{0 \ldots 0}_{\text{ 360 t.u.}} \ldots \\
g_{21} = 01 \ldots 1 1 \ldots 1 0 1 \ldots 1 1 \ldots 1
\end{array}
$$

Fig.~\ref{gun3} shows how simple addressable/rewritable memory units can be achieved in numerical studies of the BZ reaction. In this scheme the reactor becomes segregated into two zones by the formation of glider gun $e$. In the upper zone a single auxiliary spiral $a$ maintains a stable trajectory (Fig.~\ref{gun3}(a)). This spiral is a simple analog of memory as it represents past interactions in the reactor. When fragments from the glider gun enter the cell marked with a circle (Fig.~\ref{gun3}(b)), spiral $a$ is annihilated and the memory is erased (Fig.~\ref{gun3}(c)). This occurs periodically due to the complex rotation of the spiral (glider gun) causing changes in activator concentration at adjacent junctions. Therefore, once the original spiral has been erased this `memory' zone of the reactor is periodically updated. Eventually, another glider gun forms (Fig.~\ref{gun3}(d)) which provides a constant input to the `memory'. This reaction scheme shows how unstable junction transition causes a proliferation of glider guns.

\section{Glider gun-like structures in homogeneous BZ systems}
\label{homogeneous media}

The work described in the preceding sections involved the formation of glider gun-like structures in heterogeneous BZ networks with high and low excitability regions. Although these structures represent a step forward in terms of implementing a complete set of collision-based gates a heterogeneous architecture is imposed.

Our previous studies of collision-based computing in experimental and numerical studies have utilized homogeneous BZ systems~\cite{adamatzky_2004,ben_2005,adamatzky_delacycostello_2007,toth_chaos}. A homogeneous system has uniform excitability. In the case of the light-sensitive BZ reaction this is due to the projection of a uniform light intensity. If this uniform light intensity is close to the sub-excitable limit ($\mathit{\Phi}=0.0355$ model, $\mathit{\Phi}\approx5$~mW$\cdot$cm$^{-2}$ experiment) then these homogeneous BZ systems exhibit traveling unbounded wave fragments. However, constant generators of these fragments have not been identified. Wave fragments are simply formed directly from waves injected into the homogeneous sub-excitable media. These wave fragments maintain stability over relatively long distances as at this excitability level their free ends do not curl to form stable spirals.

\begin{figure}
\centering
\includegraphics[width=0.7\linewidth]{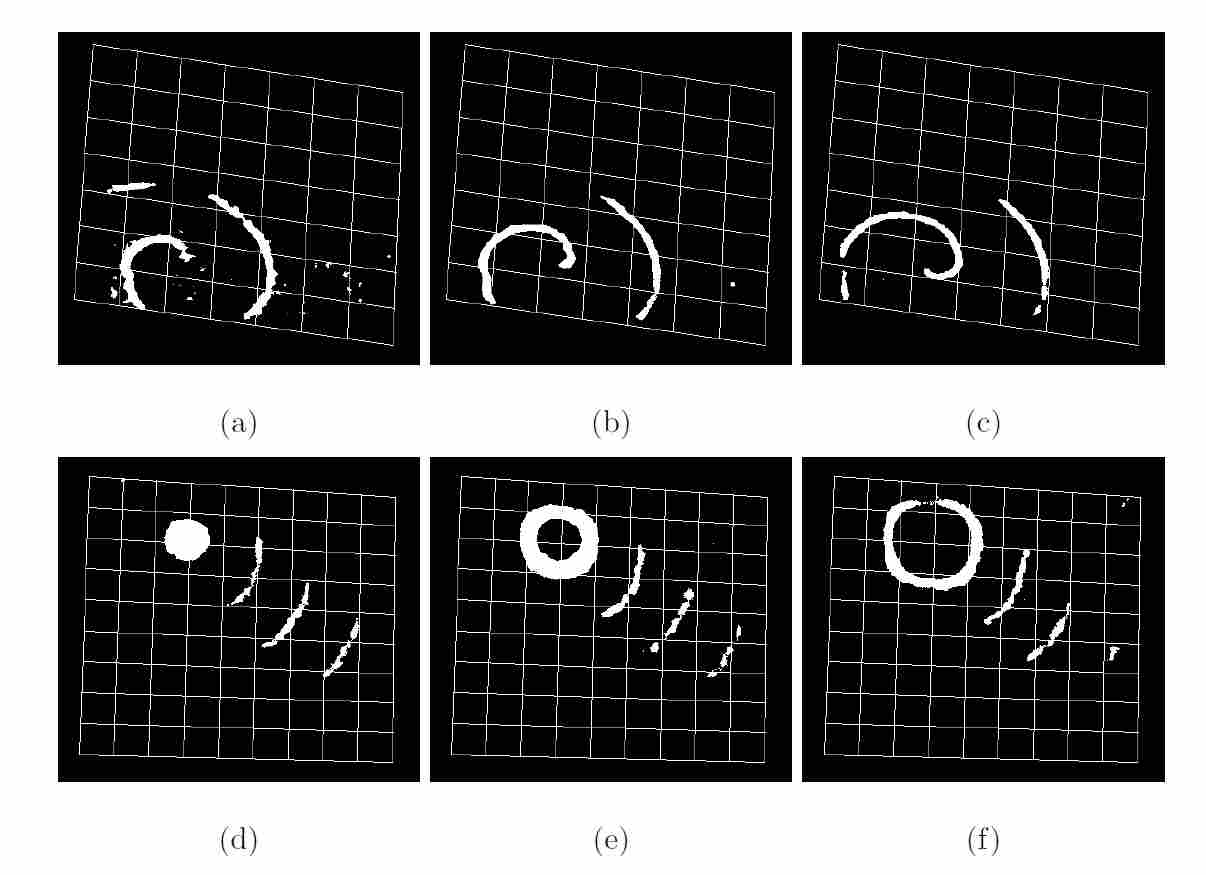}
\caption{Gliders in a homogenous sub-excitable medium. (a)-(c) snapshots of excitation dynamics with spiral initiation. (d)-(f) snapshots of
excitation dynamics with initiation via an oscillating dark region ($\mathit{\Phi}=0.034$~mW$\cdot$cm$^{-2}$). Sub-excitable light level $\mathit{\Phi}=1.5$~mW$\cdot$cm$^{-2}$. For clarity a grid has been superimposed on the figures. Grid size (a) approximately 20$\times$20mm.}
\label{spiralsub}
\end{figure}

One possible route to obtaining streams of mobile localizations in homogeneous BZ systems involves injecting fragments into a region of uniform light intensity
set at a point just above the sub-excitable limit. In experimental studies of this type, we observed the formation of an unstable spiral that underwent
spontaneous and periodic break-up. As the spiral rotated and expanded, the distance from the stable core increased and eventually the expanding arm of the spiral wave
broke in two places producing wave fragments moving in a North-East direction (Fig.~\ref{spiralsub}(a)-(c)). The continuous rotation and breakup of the spiral produced a localized
stream of wave fragments. This experimental result exhibits strong parallels with the CA model described in
Fig.~\ref{spiralhex} where the tail of an unstable rotating spiral spontaneously breaks and reforms to produce localized streams of fragments. However, this experiment involves injecting wave fragments into a weakly-excitable homogeneous medium and reliance on a highly non-linear process in the formation of an unstable spiral wave.

To successfully implement collision-based computing schemes it is necessary to have a high degree of control over the spatial evolution of excitation wave fragments. Fig.~\ref{spiralsub}(d)-(f) shows the same experimental setup with the addition of a dark region projected onto the otherwise homogeneous BZ system. Spontaneous initiation of target waves occurs in the dark region of the reactor. However, as waves move into the weakly-excitable region they spontaneously break to form a continuous stream of mobile wave fragments. This scheme imposes minimal architecture in the form of the dark area and offers benefits in terms of designing complex computing schemes as generators can easily be switched on and off. The position and firing frequencies of one or multiple generators can also be controlled.

\section{Conclusion}
\label{conclusion}

We have identified glider gun-like structures in experimental and numerical studies of the light-sensitive BZ reaction. In experimental systems with a heterogeneous network of excitable and non-excitable cells imposed, spirals rotating within excitable cells (glider guns) formed. These stable glider guns emit streams of wave fragments into all neighboring excitable cells. In numerical studies where the excitability of the network was high we observed similar stable glider gun formation. However, in numerical studies we lowered the overall excitability of the network such that certain junction transitions become unfavorable and glider guns with fewer stable streams were observed. Glider guns with intermittent firing patterns were also observed. This intermittent firing was due to the complex periodic rotation of the spiral within the excitable cell.

We showed in numerical studies how unstable glider guns could be used to implement collision-based gates via interactions with other traveling wave fragments. Therefore, theoretical ideas concerning universal computation in these systems is closer to being realized experimentally. We were able to manipulate glider streams, for example annihilate selected streams and switch periodically between two interacting streams. We were also able to show that glider guns could be formed or annihilated via specific interactions with glider streams from a second gun. We also showed examples where glider guns could be used to implement simple memory analogs. These observations highlight the potential for computing with such systems.

To truly embrace the architectureless philosophy of collision-based computing it will be necessary to identify glider gun-like structures in homogeneous BZ systems. In experiments using a weakly-excitable BZ system we identified unstable spiral waves that spontaneously break up to form continuous streams of fragments. Both this type of glider gun and the types previously identified in heterogeneous BZ systems are based on rotating spiral waves. These spiral waves periodically lose stability generating streams of excitation wave fragments. This mechanism has strong parallels with CA models based on chemical systems where gliders are formed from dynamical instabilities in rotating spiral waves.

Future work will involve a continued search for unstable glider guns in heterogeneous experimental systems. A reproducible and controllable method for forming glider gun analogs in homogeneous BZ systems will also be pursued. This will achieve the goal of having a truly architectureless computational medium with the well documented advantages that this provides. In the future, we aim to construct prototypes of arithmetical chips in the BZ medium.

\begin{acknowledgments}
This work was supported by EPSRC Grant No. GR/T11029/01.
\end{acknowledgments}

\end{document}